\documentclass[fleqn,showpacs,showkeys]{revtex4}
\usepackage{graphicx}
\usepackage{amssymb}
\usepackage{amsmath}
\usepackage{bm}
\usepackage[flushleft]{caption2}

\begin{document}
\setcounter{page}{1}
\title{Field Equations in the Complex Quaternion Spaces}
\author{Zi-Hua Weng}
\email{xmuwzh@xmu.edu.cn.}
\affiliation{School of Physics and Mechanical \& Electrical Engineering, \\Xiamen University, Xiamen 361005, China}

\begin{abstract}
The paper aims to adopt the complex quaternion and octonion to formulate the field equations for electromagnetic and gravitational fields. Applying the octonionic representation enables one single definition to combine some physics contents of two fields, which were considered to be independent of each other in the past. J. C. Maxwell applied simultaneously the vector terminology and the quaternion analysis to depict the electromagnetic theory. This method edified the paper to introduce the quaternion and octonion spaces into the field theory, in order to describe the physical feature of electromagnetic and gravitational fields, while their coordinates are able to be the complex number. The octonion space can be separated into two subspaces, the quaternion space and the S-quaternion space. In the quaternion space, it is able to infer the field potential, field strength, field source, field equations, and so forth, in the gravitational field. In the S-quaternion space, it is able to deduce the field potential, field strength, field source, and so forth, in the electromagnetic field. The results reveal that the quaternion space is appropriate to describe the gravitational features; meanwhile the S-quaternion space is proper to depict the electromagnetic features.
\end{abstract}



\maketitle

\section{\label{sec:level1}INTRODUCTION}

J. C. Maxwell described the physical feature of electromagnetic field with the vector as well as the quaternion. In 1843 W. R. Hamilton invented the quaternion, and J. T. Graves invented the octonion. Two years later, A. Cayley reinvented the octonion independently in 1845. During two decades from that time on, the scientists and engineers separated the quaternion into the scalar part and vector part, to facilitate its application in the engineering. In 1873 Maxwell mingled naturally the quaternion analysis and vector terminology to depict the electromagnetic feature in his works. Recently some scholars begin to study the physics feature of gravitational field with the algebra of quaternions.

The ordered couple of quaternions compose the octonion. On the contrary, the octonion can be separated into two parts. This case is similar to the complex number, which can be separated into the real number and the imaginary number. The octonion is able to be divided into two parts as well, the quaternion and the $S$-quaternion (short for the second quaternion), and their coordinates are able to be complex numbers. For the convenience of description, the quaternion described in the following context sometimes includes not only the quaternion but also the $S$-quaternion.

Since the vector terminology can describe the electromagnetic and gravitational theories, the quaternion should be able to depict these two theories also. In the paper, the quaternion space is suitable to describe the gravitational features, and the $S$-quaternion space is proper to depict the electromagnetic features. And it may be one approach to figure out some puzzles in the electromagnetic and gravitational theories described with the vector.

In recent years applying the quaternion to study the electromagnetic feature has been becoming one significant research orientation, and it continues the development trend of gradual deepening and expanding. Increasingly, the focus is being placed on the depiction discrepancy of electromagnetic features between the quaternion and vector. The related research into the discrepancy is coming out all the time.

Some scholars have been applying the quaternion analysis to study the electromagnetic and gravitational theories, trying to promote the further progress of these two field theories. V. Majernik \cite{majernik1} described the electromagnetic theory with the complex quaternion. W. M. Honig \cite{honig} and A. Singh \cite{singh} applied respectively the complex quaternion to deduce directly the Maxwell's equations in the classical electromagnetic theory. S. Demir \cite{demir1} etc studied the electromagnetic theory with the hyperbolic quaternion. K. Morita \cite{morita} researched the quaternion field theory. S. M. Grusky \cite{grusky} etc utilized the quaternion to investigate the time-dependent electromagnetic field. H. T. Anastassiu \cite{anastassiu} etc applied the quaternion to describe the electromagnetic feature. J. G. Winans \cite{winans} described the physics quantities with the quaternion. Meanwhile the complex quaternion has certainly piqued scholars' interest in researching the gravitational theory. J. Edmonds \cite{edmonds} utilized the quaternion to depict the wave equation and gravitational theory in the curved space-time. F. A. Doria \cite{doria} adopted the quaternion to research the gravitational theory. A. S. Rawat \cite{rawat} etc discussed the gravitational field equation with the quaternion treatment. Moreover a few scholars applied the octonion analysis to study the electromagnetic and gravitational theories. M. Gogberashvili \cite{gogberashvili} discussed the electromagnetic theory with the octonion. V. L. Mironov \cite{mironov} etc applied the octonion analysis to represent the Maxwell's equations and relevant physics features. S. Demir \cite{demir2} etc applied the algebra of octonions to study the gravitational field equations. In the paper, we shall focus on the application of the complex quaternion on the electromagnetic and gravitational theory field.

At present there are mainly three description methods for the electromagnetic theory. (1) Vector analysis. It is ripe enough the electromagnetic theory described with the three-dimensional vector. However the theory has a questionable logicality, because the deduction of Maxwell's equations is dealt with the participation of the current continuity equation. Logically, the electromagnetic strength determines the electromagnetic source via the Maxwell's equations, and then the electromagnetic source concludes the current continuity equation. On the contrary, the electromagnetic source decides the displacement current in the Maxwell's equations via the current continuity equation. This description method results in the problem of logic circulation. The puzzle urges other scholars to look for the new description method. (2) Real quaternion analysis. In the electromagnetic theory described with the real quaternion, it is able to deduce directly the electromagnetic field equations. But its displacement current's direction and the gauge condition of field potential are respectively different to that in the classical electromagnetic theory. The limited progress arouses the scholar's enthusiasm to reapply the complex quaternion to depict the electromagnetic theory. (3) Complex quaternion analysis. In the electromagnetic theory described with the complex quaternion, it is able to deduce directly the Maxwell's equations in the classical electromagnetic theory, without the help of the current continuity equation. Similarly some scholars have been applying the complex quaternion to research the gravitational theory. A comparison of the two description methods indicates that there are quite a number of sameness between the field theory described by the complex quaternion and the classical field theory described by the vector terminology, except for a little discrepancy.

In the existing studies dealt with the quaternion field theory up to now, the most of researches introduce the quaternion to rephrase simplistically the physics concept and deduction in the classical field theory. They considered the quaternion as one ordinary substitution for the complex number or the vector in the theoretical applications. This is a far cry from the expectation that an entirely new method can bring some new conclusions. In the past, the application of one new mathematic description method each time usually enlarged the range of definition of some physics concepts, bringing in new perspectives and inferences. Obviously the existing quaternion studies in the field theory so far have not achieved the expected outcome.

Making use of the comparison and analysis, it is found a few primal problems from the preceding studies. (1) The preceding studies were not able to describe simultaneously the electromagnetic and gravitational fields. These existing researches divide the electromagnetic and gravitational fields into two isolated parts, and then describe respectively the two fields. In the paper, the electromagnetic field and gravitational field can combine together to become one unitary field in the theoretical description, depicting the physics features of two fields simultaneously. (2) The preceding studies were difficult to unify similar physics quantities of two fields into the single definition. In the existing researches, the field potential (or field strength, field source) of gravitational field is different to that of electromagnetic field, and it is not able to unify them into the single definition. But the paper is able to unify the field potential (or field strength, field source) of electromagnetic and gravitational fields into the single definition.

In the paper, the author explores one new description method, introducing the complex quaternion space into the field theory, to describe the physical feature of electromagnetic and gravitational fields. This method's inferences cover the most of conclusions of electromagnetic and gravitational fields described with the vector.

\section{\label{sec:level1}Field Equations}

The octonion space $\mathbb{O}$ can be separated into two orthogonal subspaces, the quaternion space $\mathbb{H}_g$ and the $S$-quaternion space $\mathbb{H}_e$ . And the quaternion space $\mathbb{H}_g$ is independent of the $S$-quaternion space $\mathbb{H}_e$ . The quaternion space $\mathbb{H}_g$ is suitable to describe the feature of gravitational field, while the $S$-quaternion space $\mathbb{H}_e$ is proper to depict the property of electromagnetic field.

In the quaternion space $\mathbb{H}_g$ for the gravitational field, the basis vector is $\mathbb{H}_g = ( \emph{\textbf{i}}_0, \emph{\textbf{i}}_1, \emph{\textbf{i}}_2, \emph{\textbf{i}}_3 )$, the radius vector is $\mathbb{R}_g = \emph{i} r_0 \emph{\textbf{i}}_0 + \Sigma r_k \emph{\textbf{i}}_k$, and the velocity is $\mathbb{V}_g = \emph{i} v_0 \emph{\textbf{i}}_0 + \Sigma v_k \emph{\textbf{i}}_k$. The gravitational potential is $\mathbb{A}_g = \emph{i} a_0 \emph{\textbf{i}}_0 + \Sigma a_k \emph{\textbf{i}}_k$, the gravitational strength is $\mathbb{F}_g = f_0 \emph{\textbf{i}}_0 + \Sigma f_k \emph{\textbf{i}}_k$, and the gravitational source is $\mathbb{S}_g = \emph{i} s_0 \emph{\textbf{i}}_0 + \Sigma s_k \emph{\textbf{i}}_k$. In the $\emph{S}$-quaternion space $\mathbb{H}_e$ for the electromagnetic field, the basis vector is $\mathbb{H}_e = ( \emph{\textbf{I}}_0, \emph{\textbf{I}}_1, \emph{\textbf{I}}_2, \emph{\textbf{I}}_3 )$, the radius vector is $\mathbb{R}_e = \emph{i} R_0 \emph{\textbf{I}}_0 + \Sigma R_k \emph{\textbf{I}}_k$, and the velocity is $\mathbb{V}_e = \emph{i} V_0 \emph{\textbf{I}}_0 + \Sigma V_k \emph{\textbf{I}}_k$. The electromagnetic potential is $\mathbb{A}_e = \emph{i} A_0 \emph{\textbf{I}}_0 + \Sigma A_k \emph{\textbf{I}}_k$, the electromagnetic strength is $\mathbb{F}_e = F_0 \emph{\textbf{I}}_0 + \Sigma F_k \emph{\textbf{I}}_k$, and the electromagnetic source is $\mathbb{S}_e = i S_0 \emph{\textbf{I}}_0 + \Sigma S_k \emph{\textbf{I}}_k$. Herein $\mathbb{H}_e = \mathbb{H}_g \circ \emph{\textbf{I}}_0$. The symbol $\circ$ denotes the octonion multiplication. $r_j, ~v_j, ~a_j, ~s_j, ~R_j, ~V_j, ~A_j, ~S_j, ~f_0$, and $F_0$ are all real. $f_k$ and $F_k$ are the complex numbers. $i$ is the imaginary unit. $\emph{\textbf{i}}_0 = 1$. $j = 0, ~1, ~2, ~3$. $k = 1, ~2, ~3$.

These two orthogonal quaternion spaces, $\mathbb{H}_g$ and $\mathbb{H}_e$, composes one octonion space, $\mathbb{O} = \mathbb{H}_g + \mathbb{H}_e$. In the octonion space $\mathbb{O}$ for the electromagnetic and gravitational fields, the octonion radius vector is $\mathbb{R} = \mathbb{R}_g + k_{eg} \mathbb{R}_e$, the octonion velocity is $\mathbb{V} = \mathbb{V}_g + k_{eg} \mathbb{V}_e$ , with $k_{eg}$ being one coefficient. Meanwhile the octonion field potential is $\mathbb{A} = \mathbb{A}_g + k_{eg} \mathbb{A}_e$, the octonion field strength is $\mathbb{F} = \mathbb{F}_g + k_{eg} \mathbb{F}_e$. From here on out, some physics quantities are extended from the quaternion functions to the octonion functions, according to the characteristics of octonion. Apparently $\mathbb{V}$, $\mathbb{A}$, and their differential coefficients are all octonion functions of $\mathbb{R}$.

The octonion definition of field strength is,
\begin{equation}
\mathbb{F} = \square \circ \mathbb{A}  ~,
\end{equation}
where $\mathbb{F}_g = \square \circ \mathbb{A}_g$, $\mathbb{F}_e = \square \circ \mathbb{A}_e$. The gauge conditions of field potential are chosen as $f_0 = 0$ and $F_0 = 0$. The operator is $ \square = \emph{i} \emph{\textbf{i}}_0 \partial_0 + \Sigma \emph{\textbf{i}}_k \partial_k$. $ \nabla = \Sigma \emph{\textbf{i}}_k \partial_k$, $ \partial_j = \partial / \partial r_j$. $v_0$ is the speed of light.

The octonion field source $\mathbb{S}$ of the electromagnetic and gravitational fields can be defined as,
\begin{eqnarray}
\mu \mathbb{S} && = - ( \emph{i} \mathbb{F} / v_0 + \square )^* \circ  \mathbb{F}
\nonumber \\
&& = \mu_g \mathbb{S}_g + k_{eg} \mu_e \mathbb{S}_e - ( \emph{i} \mathbb{F} / v_0 )^* \circ \mathbb{F} ~,
\end{eqnarray}
where $\mu$, $\mu_g$, and $\mu_e$ are coefficients. $\mu_g < 0$, and $\mu_e > 0$. $*$ denotes the conjugation of octonion. In general, the contribution of the tiny term, $( \emph{i} \mathbb{F}^* \circ \mathbb{F}/ v_0 )$, in the above could be neglected.

According to the coefficient $k_{eg}$ and the basis vectors, $\mathbb{H}_g$ and $\mathbb{H}_e$, the octonion definition of the field source can be separated into two parts,
\begin{eqnarray}
&& \mu_g \mathbb{S}_g = - \square^* \circ \mathbb{F}_g  ~,
\\
&& \mu_e \mathbb{S}_e = - \square^* \circ \mathbb{F}_e  ~,
\end{eqnarray}
where Eq.(3) is the definition of gravitational source, while Eq.(4) is the definition of electromagnetic source.

The octonion velocity $\mathbb{V}$ is defined as,
\begin{equation}
\mathbb{V} = v_0 \partial_0 \mathbb{R}  ~,
\end{equation}
where in the case for single one particle, a comparison with the classical electromagnetic (and gravitational) theory reveals that, $\mathbb{S}_g = m \mathbb{V}_g$, and $\mathbb{S}_e = q \mathbb{V}_e$. $m$ is the mass density, while $q$ is the density of electric charge.

\begin{table}[ht]
\caption{The multiplication table of octonion.} \label{tab:table3}
\centering
\begin{tabular}{ccccccccc}
\hline
$ $ & $1$ & $\emph{\textbf{i}}_1$  &
$\emph{\textbf{i}}_2$ & $\emph{\textbf{i}}_3$  &
$\emph{\textbf{I}}_0$ & $\emph{\textbf{I}}_1$
& $\emph{\textbf{I}}_2$  & $\emph{\textbf{I}}_3$  \\
\hline $1$ & $1$ & $\emph{\textbf{i}}_1$  & $\emph{\textbf{i}}_2$ &
$\emph{\textbf{i}}_3$  & $\emph{\textbf{I}}_0$  &
$\emph{\textbf{I}}_1$
& $\emph{\textbf{I}}_2$  & $\emph{\textbf{I}}_3$  \\
$\emph{\textbf{i}}_1$ & $\emph{\textbf{i}}_1$ & $-1$ &
$\emph{\textbf{i}}_3$  & $-\emph{\textbf{i}}_2$ &
$\emph{\textbf{I}}_1$
& $-\emph{\textbf{I}}_0$ & $-\emph{\textbf{I}}_3$ & $\emph{\textbf{I}}_2$  \\
$\emph{\textbf{i}}_2$ & $\emph{\textbf{i}}_2$ &
$-\emph{\textbf{i}}_3$ & $-1$ & $\emph{\textbf{i}}_1$  &
$\emph{\textbf{I}}_2$  & $\emph{\textbf{I}}_3$
& $-\emph{\textbf{I}}_0$ & $-\emph{\textbf{I}}_1$ \\
$\emph{\textbf{i}}_3$ & $\emph{\textbf{i}}_3$ &
$\emph{\textbf{i}}_2$ & $-\emph{\textbf{i}}_1$ & $-1$ &
$\emph{\textbf{I}}_3$  & $-\emph{\textbf{I}}_2$
& $\emph{\textbf{I}}_1$  & $-\emph{\textbf{I}}_0$ \\

$\emph{\textbf{I}}_0$ & $\emph{\textbf{I}}_0$ &
$-\emph{\textbf{I}}_1$ & $-\emph{\textbf{I}}_2$ &
$-\emph{\textbf{I}}_3$ & $-1$ & $\emph{\textbf{i}}_1$
& $\emph{\textbf{i}}_2$  & $\emph{\textbf{i}}_3$  \\
$\emph{\textbf{I}}_1$ & $\emph{\textbf{I}}_1$ &
$\emph{\textbf{I}}_0$ & $-\emph{\textbf{I}}_3$ &
$\emph{\textbf{I}}_2$  & $-\emph{\textbf{i}}_1$
& $-1$ & $-\emph{\textbf{i}}_3$ & $\emph{\textbf{i}}_2$  \\
$\emph{\textbf{I}}_2$ & $\emph{\textbf{I}}_2$ &
$\emph{\textbf{I}}_3$ & $\emph{\textbf{I}}_0$  &
$-\emph{\textbf{I}}_1$ & $-\emph{\textbf{i}}_2$
& $\emph{\textbf{i}}_3$  & $-1$ & $-\emph{\textbf{i}}_1$ \\
$\emph{\textbf{I}}_3$ & $\emph{\textbf{I}}_3$ &
$-\emph{\textbf{I}}_2$ & $\emph{\textbf{I}}_1$  &
$\emph{\textbf{I}}_0$  & $-\emph{\textbf{i}}_3$
& $-\emph{\textbf{i}}_2$ & $\emph{\textbf{i}}_1$  & $-1$ \\
\hline
\end{tabular}
\end{table}

\subsection{\label{sec:level2}Gravitational field equations}

In the quaternion space $\mathbb{H}_g = ( 1, \emph{\textbf{i}}_1, \emph{\textbf{i}}_2, \emph{\textbf{i}}_3 )$, the gravitational potential is $\mathbb{A}_g = \emph{i} a_0 + \textbf{a}$, the operator is $ \square = \emph{i} \partial_0 + \nabla $, the gravitational strength is $\mathbb{F}_g = \square \circ \mathbb{A}_g$. The definition of gravitational strength can be expressed as,
\begin{eqnarray}
\mathbb{F}_g = ( - \partial_0 a_0 + \nabla \cdot \textbf{a} ) + \emph{i} ( \partial_0 \textbf{a} + \nabla a_0 ) + \nabla \times \textbf{a} ~,
\end{eqnarray}
where $\textbf{a} = \Sigma a_k \emph{\textbf{i}}_k$; $a_0 = \varphi / v_0$, with $\varphi$ being the scalar potential of gravitational field.

For the sake of convenience the paper adopts the gauge condition, $f_0 = - \partial_0 a_0 + \nabla \cdot \textbf{a} = 0$. Further the above can be written as $\textbf{f} = \emph{i} \textbf{g} / v_0 + \textbf{b}$, with $\textbf{f} = \Sigma f_k \emph{\textbf{i}}_k$. One component of gravitational strength is the gravitational acceleration, $\textbf{g} / v_0 = \partial_0 \textbf{a} + \nabla a_0$. The other is $\textbf{b} = \nabla \times \textbf{a}$, which is similar to the magnetic flux density.

In the definition of gravitational source Eq.(3), the gravitational source is, $\mathbb{S}_g = \emph{i} s_0 + \textbf{s}$. Substituting the gravitational source $\mathbb{S}_g$ and gravitational strength $\mathbb{F}_g$ into the definition of gravitational source can express the definition as,
\begin{equation}
- \mu_g ( \emph{i} s_0 + \textbf{s} ) = ( \emph{i} \partial_0 + \nabla )^* \circ ( \emph{i} \textbf{g} / v_0 + \textbf{b} ) ~,
\end{equation}
where the gravitational coefficient is $\mu_g = 1 / ( \varepsilon_g v_0^2)$, and $\varepsilon_g = - 1 / (4 \pi G)$, with $G$ being the gravitational constant. The density of linear momentum is, $\textbf{s} = \Sigma s_k \emph{\textbf{i}}_k$.

The above can be rewritten as,
\begin{eqnarray}
- \mu_g ( \emph{i} s_0 + \textbf{s} ) =  \emph{i} ( \partial_0 \textbf{b} + \nabla^* \times \textbf{g} / v_0 ) + \emph{i} ( \nabla^* \cdot \textbf{g} / v_0 )
+ ( \nabla^* \times \textbf{b} - \partial_0 \textbf{g} / v_0 ) + \nabla^* \cdot \textbf{b}  ~.
\end{eqnarray}

Comparing both sides of the equal sign in the above will yield,
\begin{eqnarray}
&& \nabla^* \cdot \textbf{b} = 0 ~,
\\
&& \partial_0 \textbf{b} + \nabla^* \times \textbf{g} / v_0 = 0 ~,
\\
&& \nabla^* \cdot \textbf{g} / v_0 = - \mu_g s_0   ~,
\\
&& \nabla^* \times \textbf{b} - \partial_0 \textbf{g} / v_0 = - \mu_g \textbf{s} ~,
\end{eqnarray}
where $s_0 = m v_0$. $\textbf{g} = \Sigma g_k \emph{\textbf{i}}_k$, $\textbf{b} = \Sigma b_k \emph{\textbf{i}}_k$.

Eqs.(9)-(12) are the gravitational field equations. Because the gravitational constant $G$ is weak and the velocity ratio $\textbf{v} / v_0$ is tiny, the gravity produced by the linear momentum $\textbf{s}$ can be ignored in general. When $\textbf{b} = 0$ and $\textbf{a} = 0$, Eq.(11) can be degenerated into the Newton's law of universal gravitation in the classical gravitational theory.

By all appearances the quaternion operator $\square$, gravitational potential $\mathbb{A}_g$, gravitational source $\mathbb{S}_g$, radius vector $\mathbb{R}_g$, and velocity $\mathbb{V}_g$ etc are restricted by the component selection in the definitions. That is, the coordinates with the basis vector $\emph{\textbf{i}}_0$ are the imaginary numbers, while the coordinates with the basis vector $\emph{\textbf{i}}_k$ are all real. In the paper they can be written as the preceding description form in the context, in order to deduce the gravitational field equations. Those restriction conditions are the indispensable components of the classical gravitational theory.

The deduction approach of the gravitational field equations can be used as a reference to be extended to that of electromagnetic field equations.

\begin{table}[ht]
\caption{The multiplication of the operator and octonion physics quantity.}
\label{tab:table2}
\centering
\begin{tabular}{ll}
\hline
definition                  &   expression~meaning                                              \\
\hline
$\nabla \cdot \textbf{a}$   &  $-(\partial_1 a_1 + \partial_2 a_2 + \partial_3 a_3)$ \\
$\nabla \times \textbf{a}$  &  $\emph{\textbf{i}}_1 ( \partial_2 a_3
                                 - \partial_3 a_2 ) + \emph{\textbf{i}}_2 ( \partial_3 a_1
                                 - \partial_1 a_3 ) + \emph{\textbf{i}}_3 ( \partial_1 a_2
                                 - \partial_2 a_1 )$                                 \\
$\nabla a_0$                &  $\emph{\textbf{i}}_1 \partial_1 a_0
                                 + \emph{\textbf{i}}_2 \partial_2 a_0
                                 + \emph{\textbf{i}}_3 \partial_3 a_0  $             \\
$\partial_0 \textbf{a}$     &  $\emph{\textbf{i}}_1 \partial_0 a_1
                                 + \emph{\textbf{i}}_2 \partial_0 a_2
                                 + \emph{\textbf{i}}_3 \partial_0 a_3  $             \\
\hline
$\nabla \cdot \textbf{A}$   &  $-(\partial_1 A_1 + \partial_2 A_2 + \partial_3 A_3) \emph{\textbf{I}}_0 $  \\
$\nabla \times \textbf{A}$  &  $-\emph{\textbf{I}}_1 ( \partial_2
                                 A_3 - \partial_3 A_2 ) - \emph{\textbf{I}}_2 ( \partial_3 A_1
                                 - \partial_1 A_3 ) - \emph{\textbf{I}}_3 ( \partial_1 A_2
                                 - \partial_2 A_1 )$                                 \\
$\nabla \circ \textbf{A}_0$ &  $\emph{\textbf{I}}_1 \partial_1 A_0
                                 + \emph{\textbf{I}}_2 \partial_2 A_0
                                 + \emph{\textbf{I}}_3 \partial_3 A_0  $             \\
$\partial_0 \textbf{A}$     &  $\emph{\textbf{I}}_1 \partial_0 A_1
                                 + \emph{\textbf{I}}_2 \partial_0 A_2
                                 + \emph{\textbf{I}}_3 \partial_0 A_3  $             \\
\hline
\end{tabular}
\end{table}

\subsection{\label{sec:level2}Electromagnetic field equations}

In the electromagnetic theory described with the complex quaternion, it is able to deduce directly the Maxwell's equations in the classical electromagnetic theory. In this approach, substituting the $S$-quaternion for the quaternion, one can obtain the same conclusions still.

In the $\emph{S}$-quaternion space $\mathbb{H}_e = ( \emph{\textbf{I}}_0, \emph{\textbf{I}}_1, \emph{\textbf{I}}_2, \emph{\textbf{I}}_3 )$, the electromagnetic potential is $\mathbb{A}_e = \emph{i} \textbf{A}_0 + \textbf{A}$, the electromagnetic strength is $\mathbb{F}_e = \square \circ \mathbb{A}_e$. The definition of electromagnetic strength can be expressed as
\begin{eqnarray}
\mathbb{F}_e = ( - \partial_0 \textbf{A}_0 + \nabla \cdot \textbf{A} ) + \emph{i} ( \partial_0 \textbf{A} + \nabla \circ \textbf{A}_0 ) + \nabla \times \textbf{A} ~,
\end{eqnarray}
where the vector potential of electromagnetic field is $\textbf{A} = \Sigma A_k \emph{\textbf{I}}_k$. $\textbf{A}_0 = A_0 \emph{\textbf{I}}_0$; $A_0 = \phi / v_0$, with $\phi$ being the scalar potential of electromagnetic field.

For the sake of convenience the paper adopts the gauge condition, $\textbf{F}_0 = - \partial_0 \textbf{A}_0 + \nabla \cdot \textbf{A} = 0$. Therefore the above can be written as $\textbf{F} = \emph{i} \textbf{E} / v_0 + \textbf{B}$. Herein the electric field intensity is $\textbf{E} / v_0 = \partial_0 \textbf{A} + \nabla \circ \textbf{A}_0$, and the magnetic flux density is $\textbf{B} = \nabla \times \textbf{A}$. $\textbf{F}_0 = F_0 \emph{\textbf{I}}_0$. $\textbf{F} = \Sigma F_k \emph{\textbf{I}}_k$.

In the definition of electromagnetic source Eq.(4), the electromagnetic source is, $\mathbb{S}_e = \emph{i} \textbf{S}_0 + \textbf{S}$. Substituting the electromagnetic source $\mathbb{S}_e$ and electromagnetic strength $\mathbb{F}_e$ into the definition of electromagnetic source can separate the definition as,
\begin{equation}
- \mu_e ( \emph{i} \textbf{S}_0 + \textbf{S} ) = ( \emph{i} \partial_0 + \nabla )^* \circ ( \emph{i} \textbf{E} / v_0 + \textbf{B} ) ~,
\end{equation}
where the electromagnetic coefficient is $\mu_e = 1 / ( \varepsilon_e v_0^2)$, with $\varepsilon_e$ being the permittivity.

The above can be rewritten as,
\begin{eqnarray}
- \mu_e ( \emph{i} \textbf{S}_0 + \textbf{S} ) = \emph{i} ( \partial_0 \textbf{B} + \nabla^* \times \textbf{E} / v_0 ) + \emph{i} ( \nabla^* \cdot \textbf{E} / v_0 )
+ ( \nabla^* \times \textbf{B} - \partial_0 \textbf{E} / v_0 ) + \nabla^* \cdot \textbf{B}  ~.
\end{eqnarray}

Comparing the variables on both sides of the equal sign in the above will reason out the Maxwell's equations as follows,
\begin{eqnarray}
&& \nabla^* \cdot \textbf{B} = 0 ~,
\\
&& \partial_0 \textbf{B} + \nabla^* \times \textbf{E} / v_0 = 0 ~,
\\
&& \nabla^* \cdot \textbf{E} / v_0 = - \mu_e \textbf{S}_0   ~,
\\
&& \nabla^* \times \textbf{B} - \partial_0 \textbf{E} / v_0 = - \mu_e \textbf{S} ~,
\end{eqnarray}
where $\textbf{S}_0 = S_0 \emph{\textbf{I}}_0$, $S_0 = q V_0$. $\textbf{E} = \Sigma E_k \emph{\textbf{I}}_k$, $\textbf{B} = \Sigma B_k \emph{\textbf{I}}_k$. The density of electric current is $\textbf{S} = \Sigma S_k \emph{\textbf{I}}_k$. For the charged particles, there may be, $\mathbb{V}_e = \mathbb{V}_g \circ \textbf{I} ( \emph{\textbf{I}}_j )$. And the unit $\textbf{I} ( \emph{\textbf{I}}_j )$ is one function of $\emph{\textbf{I}}_j$, with $\textbf{I} ( \emph{\textbf{I}}_j )^* \circ \textbf{I} ( \emph{\textbf{I}}_j ) = 1$.

In a similar way the electromagnetic potential $\mathbb{A}_e$, electromagnetic source $\mathbb{S}_e$, radius vector $\mathbb{R}_e$ , and velocity $\mathbb{V}_e$ etc are restricted by the component selection of the definitions. That is, the coordinates with the basis vector $\emph{\textbf{I}}_0$ are the imaginary numbers, while the coordinates with the basis vector $\emph{\textbf{I}}_k$ are all real. In the paper they can be written as the preceding description form in the context, in order to deduce the Maxwell's equations. Those restriction conditions are the indispensable components of the classical electromagnetic theory.

On the analogy of the coordinate definition of complex coordinate system, one can define the coordinate of octonion, which involves the quaternion and $\emph{S}$-quaternion simultaneously. In the octonion coordinate system, the octonion physics quantity can be defined as $ \{ ( \emph{i} c_0 + \emph{i} d_0 \textbf{\emph{I}}_0 ) \circ \emph{\textbf{i}}_0 + \Sigma ( c_k + d_k \textbf{\emph{I}}_0^* ) \circ \emph{\textbf{i}}_k \} $. It means that there are the quaternion coordinate $c_k$ and the $S$-quaternion coordinate $d_k \emph{\textbf{I}}_0^*$ for the basis vector $\emph{\textbf{i}}_k$, while the quaternion coordinate $c_0$ and the $S$-quaternion coordinate $d_0 \emph{\textbf{I}}_0$ for the basis vector $\emph{\textbf{i}}_0$. Herein $c_j$ and $d_j$ are all real.

\section{\label{sec:level1}Equivalent Transformation}

\subsection{\label{sec:level2}Gravitational Field Equations}

Making use of the transforming of the basis vectors of $\nabla$ in the quaternion operator $\square$, it is able to translate further the gravitational field equations, Eqs.(9)-(12), from the quaternion space $\mathbb{H}_g$ into that in the three-dimensional vector space. In the three-dimensional vector space $(\emph{\textbf{j}}_1 ,~\emph{\textbf{j}}_2 ,~\emph{\textbf{j}}_3)$, the operator is $\blacktriangledown = \Sigma \emph{\textbf{j}}_k \partial_k$, with $\emph{\textbf{j}}_k^{~2} = 1$.

In the gravitational field equations, the operator $\nabla$ should be substituted by the operator $\blacktriangledown$. Meanwhile $\textbf{b}$, $\textbf{g}$, and $\textbf{s}$ are substituted by $\textbf{b}^\prime = \Sigma b_k \emph{\textbf{j}}_k$, $\textbf{g}^\prime = \Sigma g_k \emph{\textbf{j}}_k$, and $\textbf{s}^\prime = \Sigma s_k \emph{\textbf{j}}_k$ respectively. The cross and dot products of quaternion are substituted by that of vector respectively. And the gravitational field equations, Eqs.(9)-(12), can be transformed into,
\begin{eqnarray}
&& - \blacktriangledown \cdot \textbf{b}^\prime = 0 ~,
\\
&& \partial_0 \textbf{b}^\prime - \blacktriangledown \times \textbf{g}^\prime / v_0 = 0 ~,
\\
&& \blacktriangledown \cdot \textbf{g}^\prime / v_0 = - \mu_g s_0   ~,
\\
&& \blacktriangledown \times \textbf{b}^\prime + \partial_0 \textbf{g}^\prime / v_0 = \mu_g \textbf{s}^\prime ~,
\end{eqnarray}
where $\textbf{b}^\prime = \blacktriangledown \times \textbf{a}^\prime$, $\textbf{g}^\prime / v_0 = \partial_0 \textbf{a}^\prime + \blacktriangledown a_0$. $\textbf{a}^\prime = \Sigma a_k \emph{\textbf{j}}_k$.

Expressing the above into the scalar equations will reveal that Eqs.(9)-(12) and Eqs.(20)-(23) are equivalent to each other. And the definition of Eq.(6) requires to substitute $\textbf{g}^\prime$ by $\textbf{g}^{\prime\prime} = - \textbf{g}^\prime$, in order to approximate to the Newton's law of universal gravitation in the classical gravitational theory as near as possible. Therefore the gravitational field equations, Eqs.(20)-(23), are able to be transformed into that in the three-dimensional vector space,
\begin{eqnarray}
&& \blacktriangledown \cdot \textbf{b}^\prime = 0 ~,
\\
&& \partial_0 \textbf{b}^\prime + \blacktriangledown \times \textbf{g}^{\prime\prime} / v_0 = 0 ~,
\\
&& \blacktriangledown \cdot \textbf{g}^{\prime\prime} / v_0 = \mu_g s_0   ~,
\\
&& \blacktriangledown \times \textbf{b}^\prime - \partial_0 \textbf{g}^{\prime\prime} / v_0 = \mu_g \textbf{s}^\prime ~.
\end{eqnarray}

Comparing Eq.(26) with Poisson equation for the Newton's law of universal gravitation reveals that those two equations are the same formally. In case the gravitational strength component $\textbf{b}$ is weak and the velocity ratio $\textbf{v} / v_0$ is quite tiny, Eq.(26) will be reduced into the Newton's law of universal gravitation in the classical gravitational theory.

\subsection{\label{sec:level2}Electromagnetic field equations}

By means of the transformation of basis vectors of $\nabla$ in the quaternion operator, it is able to translate further the electromagnetic field equations, Eqs.(16)-(19), from the $\emph{S}$-quaternion space $\mathbb{H}_e$ into that in the three-dimensional vector space $(\emph{\textbf{j}}_1 ,~\emph{\textbf{j}}_2 ,~\emph{\textbf{j}}_3)$. In the electromagnetic field equations, the operator $\nabla$ is substituted by $\blacktriangledown$. And $\textbf{B}$, $\textbf{E}$, and $\textbf{S}$ are substituted by $\textbf{B}^\prime$, $\textbf{E}^\prime$, and $\textbf{S}^\prime$ respectively. Therefore the electromagnetic field equations, Eqs.(16)-(19), can be transformed into,
\begin{eqnarray}
&& \blacktriangledown \cdot \textbf{B}^\prime = 0 ~,
\\
&& \partial_0 \textbf{B}^\prime + \blacktriangledown \times \textbf{E}^\prime / v_0 = 0 ~,
\\
&& \blacktriangledown \cdot \textbf{E}^\prime / v_0 = - \mu_e S_0   ~,
\\
&& \blacktriangledown \times \textbf{B}^\prime - \partial_0 \textbf{E}^\prime / v_0 = - \mu_e \textbf{S}^\prime ~,
\end{eqnarray}
where $\textbf{B}^\prime = - \blacktriangledown \times \textbf{A}^\prime$, $\textbf{E}^\prime  / v_0 = \partial_0 \textbf{A}^\prime  + \blacktriangledown A_0$. $\textbf{A}^\prime  = \Sigma A_k \emph{\textbf{j}}_k$, $\textbf{S}^\prime  = \Sigma S_k \emph{\textbf{j}}_k$.

Expressing the above into the scalar equations will reveal that Eqs.(16)-(19) and Eqs.(28)-(31) are equivalent to each other. And the definition of Eq.(13) requires to substitute $\textbf{E}^\prime$ and $\textbf{B}^\prime$ by $\textbf{E}^{\prime\prime} = - \textbf{E}^\prime$ and $\textbf{B}^{\prime\prime} = - \textbf{B}^\prime¡ä$ respectively, in order to approximate to the Maxwell's equations in the classical electromagnetic theory as near as possible. Therefore the electromagnetic field equations, Eqs.(28)-(31), are able to be transformed into that in the three-dimensional vector space,
\begin{eqnarray}
&& \blacktriangledown \cdot \textbf{B}^{\prime\prime} = 0 ~,
\\
&& \partial_0 \textbf{B}^{\prime\prime} + \blacktriangledown \times \textbf{E}^{\prime\prime} / v_0 = 0 ~,
\\
&& \blacktriangledown \cdot \textbf{E}^{\prime\prime} / v_0 = \mu_e S_0   ~,
\\
&& \blacktriangledown \times \textbf{B}^{\prime\prime} - \partial_0 \textbf{E}^{\prime\prime} / v_0 = \mu_e \textbf{S}^\prime ~.
\end{eqnarray}

Comparing Eqs.(32)-(35) with the Maxwell's equations in the classical electromagnetic theory states that those two field equations are equivalent to each other. It means that in the electromagnetic theory described with the quaternion, without the participation of the current continuity equation, it is still able to reason out the Maxwell's equations in the classical electromagnetic theory.

\section{\label{sec:level1}Conclusions and Discussions}

Applying the complex quaternion space, the paper is able to describe simultaneously the electromagnetic field equations and the gravitational field equations. The spaces of gravitational and electromagnetic fields both can be chosen as the quaternion spaces. Furthermore some coordinates of those two quaternion spaces and relevant physics quantities may be the imaginary numbers.

In the definitions of the quaternion operator, field potential, field source, radius vector, and velocity etc, the coordinates with the basis vector $\emph{\textbf{I}}_0$ or $\emph{\textbf{i}}_0$ are all imaginary numbers. However this simple case is still able to result in many components of other physics quantities to become the imaginary numbers even the complex numbers. In the definitions, only by means of the confinement of the component selection can we achieve the quaternion field theory which approximating to the classical field theory, enabling it to cover the classical field theory. Therefore it is able to be said that the component selection are the essential ingredients for the field theory too.

In the quaternion space, it is able to deduce the gravitational field equations as well as the gauge condition of gravitational potential etc. The Newton's law of universal gravitation in the classical gravitational theory can be derived from the gravitational field equations described with the quaternion. Two components of the gravitational strength are the counterpart of the linear acceleration and of the precession angular velocity. In general, the component of the gravitational strength, which is corresponding to the precession angular velocity of the gyroscopic torque dealt with the angular momentum, is comparatively tiny.

In the $S$-quaternion space, it is able to reason out directly the electromagnetic field equations as well as the gauge condition of electromagnetic potential etc. The electromagnetic field equations described with the $S$-quaternion can be equivalently translated into the Maxwell's equations of the classical electromagnetic theory in the three-dimensional vector space. Moreover the definition of electromagnetic strength and the gauge condition of electromagnetic potential both can be equivalently translated respectively into that in the classical electromagnetic theory.

It should be noted that the paper discussed only the electromagnetic field equations and the gravitational field equations etc described with the complex quaternion, and the equivalent transformation of those field equations into that in the three-dimensional vector space. However it clearly states that the introducing of the complex quaternion space is able to availably describe the physics features of electromagnetic and gravitational fields. This will afford the theoretical basis for the further relevant theoretical analysis, and is helpful to apply the complex quaternion analysis to study the angular momentum, electric/magnetic dipole moment, torque, energy, and force etc in the electromagnetic and gravitational fields.

\begin{acknowledgments}
The author is indebted to the anonymous referee for their valuable and constructive comments on the previous manuscript. This project was supported partially by the National Natural Science Foundation of China under grant number 60677039.
\end{acknowledgments}

\end{document}